\documentclass[aps,preprint,showpacs,preprintnumbers,amsmath,amssymb]{revtex4}
\usepackage{amsmath,mathrsfs,amsbsy,color,graphicx,bm,amsthm,amsfonts}
\usepackage{units}
\usepackage{bbm}
\usepackage{times}
\usepackage{dcolumn}
\usepackage{mathrsfs}
\usepackage{amsmath,amssymb,epsfig}
\usepackage{amsmath}
\newcommand{\udots}{\mathinner{\mskip1mu\raise1pt\vbox{\kern7pt\hbox{.}}
\mskip2mu\raise4pt\hbox{.}\mskip2mu\raise7pt\hbox{.}\mskip1mu}}
\begin{document}
\title{Genuine tripartite nonlocality and  entanglement in curved spacetime }
\author{Shu-Min Wu$^1$, Hao-Sheng Zeng$^2$\footnote{Email: hszeng@hunnu.edu.cn}}
\affiliation{$^1$ Department of Physics, Liaoning Normal University, Dalian 116029, China\\
$^2$ Department of Physics, Hunan Normal University, Changsha 410081, China}


\begin{abstract}
We study the genuine tripartite nonlocality (GTN) and the genuine tripartite entanglement (GTE) of Dirac fields in the background of a Schwarzschild black hole. We find that the Hawking radiation degrades both the physically accessible GTN and the physically accessible GTE. The former suffers from ``sudden death" at some critical Hawking temperature, and the latter approaches to the nonzero asymptotic value in the limit of infinite Hawking temperature.
We also find that the Hawking effect cannot generate the physically inaccessible GTN, but can generate the physically inaccessible GTE for fermion fields in curved spacetime. These results show that on the one hand the GTN cannot pass through the event horizon of black hole, but the GTE do can, and on the other hand the surviving physically accessible GTE and the generated physically inaccessible GTE for fermions in curved spacetime are all not nonlocal. Some monogamy relations between the physically accessible GTE and the physically inaccessible GTE are found.
\end{abstract}

\vspace*{0.5cm}
\maketitle
\section{Introduction}
The concept of quantum nonlocality was firstly proposed by Einstein, Podolsky and Rosen in 1935 in their famous EPR paradox \cite{EPR1935}. Afterwards, Bell established the so-called Bell inequality for conveniently judging the existence of quantum nonlocality \cite{L1}. Two quantum systems that admit Bell inequality cannot be regarded as nonlocal, even if they are far apart in space. Contrarily, when the Bell inequality is violated, we say that the two quantum systems are inseparable or have the quantum nonlocality. Quantum nonlocality is a kind of quantum behaviors, which denies the local hidden variable (LHV) model \cite{L2,L3,L4,L5,L6,L7,L8}. Quantum nonlocality is a very important quantum resource and has been applied in different fields, such as device-independent  quantum computation, communication complexity, quantum cryptography and randomness generation \cite{L9,L10,L11,L12,L13,L14}. In the case of tripartite quantum systems, a more valuable kind of resource is involved which is called GTN.
Svetlichny introduced a kind of GTN and found the so-called Svetlichny inequality to detect its existence \cite{L15}.
Quantum nonlocality and entanglement are inextricably linked. Quantum nonlocality originates from quantum entanglement, but quantum entanglement does not imply quantum nonlocality.

The combination of quantum information science, relativity theory and quantum field theory gives us a deeper understanding of quantum mechanics. It is necessary to understand quantum effects in the relativistic framework, because the world is essentially noninertial or/and curved.
Recently, quantum entanglement under relativistic settings received much attention \cite{L16,L17,L18,L19,L20,L21,L22,L23,L24,L25,L26,L27,zhx2,zhx3}, including the bipartite entanglement for boson fields \cite{L16} and fermi fields \cite{L17} in the noninertial frames, and the bipartite entanglement in curved spacetime \cite{L18,L19,L20,L21,L22,L23,L24,L25,L26,L27,zhx2}. Bipartite entanglement under the joint influence of both environmental noise and Unruh effect \cite{L28,L29} was also studied.
Besides bipartite systems, tripartite entanglement in the relativistic framework was also investigated \cite{L30,L31,L32,L34,L35,L36,L37}. Note that, in these works, the measure of GTE is based on the concept of logarithmic negativity \cite{L38,L39}, and the measure of GTE based on concurrence \cite{L44,L45} was not used in Schwarzschild spacetime.

In this work, we study the properties of GTN and GTE of Dirac fields in the background of a Schwarzschild black hole. Assume that Alice, Bob and Charlie initially share a Greenberger-Horne-Zeilinger-like state. Alice is a Kruskal observer who stays stationarily at an asymptotically flat region, while Bob and Charlie are Schwarzschild observers who hover near the event horizon of the black hole. In addition, there are two imagined observers, anti-Bob and anti-Charlie, in the interior of the event horizon. The Hawking effect would make the information tunnelling from the exterior to the interior of the event horizon, such that a correlated state that involves the above five observers is established. Since the interior of the black hole is causally disconnected
from the exterior and observers cannot access to the interior of the event horizon, we thus call the information (including GTN and GTE) that distributes completely in the outside of the event horizon the physically accessible. Otherwise, it is called the physically inaccessible.
The main end of this paper is to study the influence of Hawking effect on the physically accessible GTN and GTE, the production of the physically inaccessible GTN and GTE, and the monogamy relationship between the physically accessible and inaccessible information.

The paper is organized as follows. In Sec. II, we briefly recall the measures of GTN and GTE for tripartite quantum systems. In Sec. III, we introduce the quantization of Dirac fields in the background of Schwarzschild black hole.  Sec. IV is the main contribution of our work, where the decay of the physically accessible information and the production of the physically inaccessible information, as well as the monogamy relationship between them, under the influence of Hawking effect, are studied. Finally, the last section is devoted to the conclusion.

\section{Measures of GTN and GTE \label{GSCDGE}}

\subsection{Measure of GTN \label{GSCDGE}}
Firstly, we briefly review the concept about GTN.
Nonlocality in tripartite systems has been considered as the manifestation of genuine tripartite correlations. Generally, local tripartite correlations shared by Alice, Bob and Charlie can be written as
\begin{eqnarray}\label{S1}
P(a,b,c|x,y,z)=\sum_\lambda p_\lambda P_\lambda(a|x)P_\lambda(b|y)P_\lambda(c|z),
\end{eqnarray}
where $\{\lambda; p_{\lambda}\}$ is the probability distribution of some hidden variable that controls the outputs $a, b, c\in\{0,1\}$ of the local measurements performed by Alice, Bob and Charlie on the
two-valued variables $x, y, z\in\{0,1\}$, and $P_\lambda(a|x)$ is the conditional probability for obtaining the output $a$ when the measurement setting is $x$ and $\lambda$. If the tripartite correlation cannot be written as the form of Eq.(\ref{S1}), then the tripartite system is said to have GTN.

In 1987, Svetlichny proposed a hybrid local-nonlocal form of correlation to measure GTN \cite{L15}. A tripartite correlation under the definition of Svetlichny is called
to be local if it admits the following local LHV model
\begin{eqnarray}\label{S2}
P(a,b,c|x,y,z)&=&\sum_\lambda p_\lambda P_\lambda(a|x)P_\lambda(b,c|y,z)+\sum_\mu p_\mu P_\mu(b|y)P_u(a,c|x,z) \\ \nonumber
&+&\sum_\nu p_\nu P_\lambda(c|z)P_\nu(a,b|x,y),
\end{eqnarray}
where $\sum_\lambda p_\lambda+\sum_\mu p_\mu+\sum_\nu p_\nu=1$.  This kind of correlation is regarded as Svetlichny local, otherwise it is  Svetlichny nonlocal \cite{L40,L41,L42}.

Assuming that Alice, Bob and Charlie share a state $\rho$ of some three-qubit system. Alice performs measurements on the observable $\boldsymbol{A}=\boldsymbol{a}\cdot\boldsymbol{\sigma}$ and $\boldsymbol{A'}=\boldsymbol{a'}\cdot\boldsymbol{\sigma}$, where $\boldsymbol{a}=(a_1, a_2, a_3)$, $\boldsymbol{a'}=(a'_1, a'_2, a'_3)\in \mathbb{R}^3$ are any three-dimensional unit vectors, and $\boldsymbol{\sigma}=(\sigma_1, \sigma_2, \sigma_3)$ is the vector of Pauli matrices. Similar measurements are performed by Bob and Charlie, respectively, on the observable $\boldsymbol{B}$, $\boldsymbol{B'}$ and observable $\boldsymbol{C}$, $\boldsymbol{C'}$.
Introducing Svetlichny operator
$$S=(\boldsymbol{A}+\boldsymbol{A'})\otimes(\boldsymbol{B}\otimes\boldsymbol{C'}+\boldsymbol{B'}\otimes\boldsymbol{C})
+(\boldsymbol{A}-\boldsymbol{A'})\otimes(\boldsymbol{B}\otimes\boldsymbol{C}-\boldsymbol{B'}\otimes\boldsymbol{C'}),$$
then the Svetlichny inequality
\begin{eqnarray}\label{S3}
{\rm tr}(S\rho)\leq4
\end{eqnarray}
fulfills for any state $\rho$ that admits LHV model,  where $S$ is taken over all possible Svetlichny operators, i.e., the measurements are taken over all the directions of three-dimensional space.
Equivalently, if a tripartite state violates the Svetlichny inequality for some
Svetlichny operator, then this state is genuinely nonlocal.
For convenience, we introduce the Svetlichny value, i.e., the maximum expectation value of
Svetlichny operator\cite{L7,L8,L43}
\begin{eqnarray}\label{S4}
S(\rho)=\max_S {\rm tr}(S\rho)
\end{eqnarray}
to detect the GTN of a given three-qubit state $\rho$.
For the three-qubit $X$ state with density matrix
\begin{eqnarray}\label{S5}
 \rho_X= \left(\!\!\begin{array}{cccccccc}
n_{1} & 0 & 0 & 0 & 0 & 0 & 0 & c_{1}\\
0 & n_{2} &0  &0  & 0 & 0 & c_{2} & 0\\
0 & 0 & n_{3} & 0 & 0 &  c_{3} &0  & 0\\
0 & 0 & 0 & n_{4} & c_{4} & 0 & 0 &0 \\
0 & 0 & 0 & c^*_{4} & m_{4} &0 & 0 &0 \\
0 & 0 & c^*_{3} & 0 & 0 & m_{3} & 0 & 0\\
0 & c^*_{2} & 0 & 0 & 0 & 0 & m_{2} &0 \\
c^*_{1} & 0 & 0 & 0 & 0 & 0 & 0 & m_{1}
\end{array}\!\!\right)
\end{eqnarray}
in the orthogonal basis $\{|0,0,0\rangle,|0,0,1\rangle,...,|1,1,1\rangle\}$, the Svetlichny value can be simply given by\cite{L8}
\begin{eqnarray}\label{S6}
S(\rho_X)=\max \{8\sqrt{2}|c_{i}|,4|N|\},
\end{eqnarray}
where $N=n_{1}-n_{2}-n_{3}+n_{4}-m_{4}+m_{3}+m_{2}-m_{1}$.

\subsection{Measure of GTE \label{GSCDGE}}
GTE can be defined by its opposite of biseparability.
We call a tripartite pure state $|\Psi\rangle$ is biseparable, if it has a bipartition of the form $|\Psi\rangle=|\Psi_A\rangle\otimes|\Psi_B\rangle$, where $|\Psi_A\rangle$ and $|\Psi_B\rangle$ are the monomeric or bipartite pure states. Obviously, a biseparable pure state has at least one
pure marginal. If the tripartite state $|\Psi\rangle$ is not biseparable with respect to any of its bipartition, then it is called GTE.
One can then define the so-called genuine tripartite concurrence
$C(|\Psi\rangle)=\min_{\chi_i\in\chi}\sqrt{2[1-\text{Tr}(\rho^2_{A_{\chi_i}})]}$
for describing the degrees of the GTE for the pure state $|\Psi\rangle$, where $\chi=\{A_i|B_i\}$ denotes the set of all possible bipartitions of the tripartite system, and $\rho_{A_{\chi_i}}$ is the reduced density operator of system $A$ corresponding to the bipartition $\chi_i$ \cite{L44,L33}. The GTE for a mixed state $\rho$ can be obtained by a convex roof construction
\begin{eqnarray}\label{S7}
C(\rho)=\inf_{\{p_i,|\Psi_i\rangle\}}\sum_ip_iC(|\Psi_i\rangle),
\end{eqnarray}
where the infimum takes over all possible decompositions $\rho=\sum_ip_i|\Psi_i\rangle\langle\Psi_i|$.
For three-qubit $X$ states given by Eq.(\ref{S5}), the GTE is given by
\begin{eqnarray}\label{S8}
C(\rho_X)=2\max \{0,|c_i|-\nu_i \},   i=1,\ldots,4,
\end{eqnarray}
where $\nu_i=\sum_{j\neq i}^4\sqrt{n_jm_j}$ \cite{L45}.

\section{Quantization of Dirac fields in a Schwarzschild black hole  \label{GSCDGE}}
The  Dirac equation under a general background
spacetime can be written as \cite{L46}
\begin{equation}\label{S9}
[\gamma^a e_a{}^\mu(\partial_\mu+\Gamma_\mu)]\Phi=0,
\end{equation}
where $\gamma^a$ are the Dirac matrices, the four-vectors $e_a{}^\mu$
is the inverse of the tetrad $e^a{}_\mu$, and $\Gamma_\mu=
\frac{1}{8}[\gamma^a,\gamma^b]e_a{}^\nu e_{b\nu;\mu}$ are the spin
connection coefficients. The metric of the Schwarzschild black hole can be written as
\begin{eqnarray}\label{S10}
ds^2&=&-(1-\frac{2M}{r}) dt^2+(1-\frac{2M}{r})^{-1} dr^2\nonumber\\&&+r^2(d\theta^2
+\sin^2\theta d\varphi^2),
\end{eqnarray}
where $M$ denotes the mass of the black hole. For simplicity, we take $\hbar, G, c$ and $k$ as unity in this paper.

Solving the Dirac equation of Eq.(\ref{S9}) near the event horizon of black hole, a set of positive frequency outgoing solutions for the inside and outside regions of the event horizon  can be obtained \cite{L19,L23,L47}
\begin{eqnarray}\label{S11}
\Phi^+_{{k},{\rm in}}\sim \phi(r) e^{i\omega u},
\end{eqnarray}
\begin{eqnarray}\label{S12}
\Phi^+_{{k},{\rm out}}\sim \phi(r) e^{-i\omega u},
\end{eqnarray}
where $\phi(r)$ denotes four-component Dirac spinor, $\omega$ is a monochromatic frequency, and
$u=t-r_{*}$ with the tortoise coordinate $r_{*}=r+2M\ln\frac{r-2M}{2M}$.
The modes $\Phi^+_{{k},{\rm in}}$ and $\Phi^+_{{k},{\rm out}}$ are usually called Schwarzschild
modes. According to future-directed timelike Killing vector under  each
region, particles and antiparticles will be classified.

Making an analytic continuation for Eqs.(\ref{S11}) and (\ref{S12}) according to suggestion of Damour and Ruffini, we obtain a complete
basis of positive energy modes,  i.e., the Kruskal modes \cite{L48}. Then, we can use Schwarzschild mode and Kruskal mode to expand the Dirac field, respectively, leading to the Bogoliubov transformations  between annihilation operator and creation operator under the Schwarzschild and Kruskal coordinates \cite{L49,zhx1}. After properly normalizing the state vector,  the vacuum state and excited state of the Kruskal particle in the single-mode approximation are given by
\begin{eqnarray}\label{S13}
\nonumber |0\rangle_K&=&(e^{-\frac{\omega}{T}}+1)^{-\frac{1}{2}}|0\rangle_I |0\rangle_{II}+(e^{\frac{\omega}{T}}+1)^{-\frac{1}{2}}|1\rangle_I |1\rangle_{II},\\
|1\rangle_K&=&|1\rangle_I |0\rangle_{II},
\end{eqnarray}
where $T=\frac{1}{8\pi M}$ is the Hawking temperature, $\{|n\rangle_I\}$ and $\{|n\rangle_{II}\}$ are the number states for the particle outside the region and the antiparticle inside the region of the event horizon, respectively.

\section{Evolution of the GTN and GTE in Schwarzschild black hole  \label{GSCDGE}}
Consider a Greenberger-Horne-Zeilinger-like (GHZ-like) state of the Dirac fields shared by Alice, Bob and  Charlie in the asymptotically flat region
\begin{eqnarray}\label{S14}
|\Psi\rangle_{ABC}=\alpha|0_{A}0_{B}0_{C}\rangle+\sqrt{1-\alpha^2}|1_{A}1_{B}1_{C}\rangle,
\end{eqnarray}
where $\alpha$ is the state parameter that runs from $0$ to $1$.
Now, we assume that Alice still stays at
an asymptotically flat region, while Bob and Charlie hover outside the event horizon of the black hole, then we can rewrite Eq.(\ref{S14}) in terms of Kruskal modes for Alice and Schwarzschild modes for Bob and Charlie as
\begin{eqnarray}\label{S15}
\nonumber\Psi_{AB_IB_{II}C_IC_{II}} &=& \alpha (e^{\frac{\omega}{T}}+1)^{-1} |0\rangle_A |1\rangle_{B_I}|1\rangle_{B_{II}} |1\rangle_{C_I}|1\rangle_{C_{II}}\\
\nonumber&+&\sqrt{1-\alpha^2} |1\rangle_A |1\rangle_{B_I}|0\rangle_{B_{II}} |1\rangle_{C_I}|0\rangle_{C_{II}}\\
\nonumber&+&\alpha (e^{\frac{\omega}{T}}+e^{-\frac{\omega}{T}}+2)^{-\frac{1}{2}}( |0\rangle_A |0\rangle_{B_I}|0\rangle_{B_{II}} \nonumber|1\rangle_{C_I}|1\rangle_{C_{II}}\\
\nonumber&+&|0\rangle_A |1\rangle_{B_I}|1\rangle_{B_{II}} |0\rangle_{C_I}|0\rangle_{C_{II}})\\
&+&\alpha (e^{-\frac{\omega}{T}}+1)^{-1} |0\rangle_A |0\rangle_{B_I}|0\rangle_{B_{II}} |0\rangle_{C_I}|0\rangle_{C_{II}}.
\end{eqnarray}
Generally, this is a 5-partite entangled state consisted by subsystems: subsystem $A$ observed by Alice,
subsystems $B_{I}$ and $C_{I}$ observed by Bob and Charlie outside the event horizon
of black hole, and subsystems $B_{II}$ and $C_{II}$ observed by anti-Bob and anti-Charlie inside the event horizon, respectively.

Since the interior region of black hole is causally disconnected
from the exterior region, and Alice, Bob and  Charlie cannot access the modes inside the event horizon, we thus call the modes $B_{I}$ and $C_{I}$ outside the event horizon the accessible modes, and the modes $B_{II}$ and $C_{II}$ inside the event horizon the inaccessible modes. Taking trace over the inaccessible modes on state $\Psi_{AB_IB_{II}C_IC_{II}}$, we obtain the reduced density operator $\rho_{AB_IC_I}$
\begin{eqnarray}\label{S16}
 \rho_{AB_IC_I}= \left(\!\!\begin{array}{cccccccc}
n_{1} & 0 & 0 & 0 & 0 & 0 & 0 & c_1\\
0 & n_{2} &0  &0  & 0 & 0 & 0 & 0\\
0 & 0 & n_{3} & 0 & 0 &  0 &0  & 0\\
0 & 0 & 0 & n_{4} & 0 & 0 & 0 &0 \\
0 & 0 & 0 & 0 & 0 &0 & 0 &0 \\
0 & 0 & 0 & 0 & 0 & 0 & 0 & 0\\
0 & 0 & 0 & 0 & 0 & 0 & 0 &0 \\
c_{1} & 0 & 0 & 0 & 0 & 0 & 0 & m_{1}
\end{array}\!\!\right),
\end{eqnarray}
where the matrix elements are written by
\begin{eqnarray}
\nonumber n_1&=&\alpha^2(e^{-\frac{\omega}{T}}+1)^{-2}, n_2=n_3=\alpha^2 (e^{\frac{\omega}{T}}+e^{-\frac{\omega}{T}}+2)^{-1}\\
a_4&=&\alpha^2(e^{\frac{\omega}{T}}+1)^{-2}, m_1=1-\alpha^2, c_1=\alpha\sqrt{1-\alpha^2}(e^{-\frac{\omega}{T}}+1)^{-1} \nonumber.
\end{eqnarray}
According to Eqs.(\ref{S6}) and (\ref{S8}), we obtain the Svetlichny value and GTE for state $\rho_{AB_IC_I}$,
\begin{eqnarray}\label{S17}
S(\rho_{AB_IC_I})=\max \{8\alpha\sqrt{2(1-\alpha^2)}(e^{-\frac{\omega}{T}}+1)^{-1},
4|\alpha^2(e^{\frac{\omega}{T}}+1)^{-2}(e^{\frac{\omega}{T}}-1)^{2}+\alpha^2-1|\},
\end{eqnarray}
and
\begin{eqnarray}\label{S18}
C(\rho_{AB_IC_I})=2\alpha\sqrt{1-\alpha^2}(e^{-\frac{\omega}{T}}+1)^{-1},
\end{eqnarray}
respectively. Obviously, the Svetlichny value and GTE depend not only on the state parameter $\alpha$, but also on the
Hawking temperature $T$, meaning that the
Hawking radiation of black hole will affect the physically accessible GTN and GTE between Alice, Bob and  Charlie.

\begin{figure}
\begin{minipage}[t]{0.5\linewidth}
\centering
\includegraphics[width=3.0in,height=5.2cm]{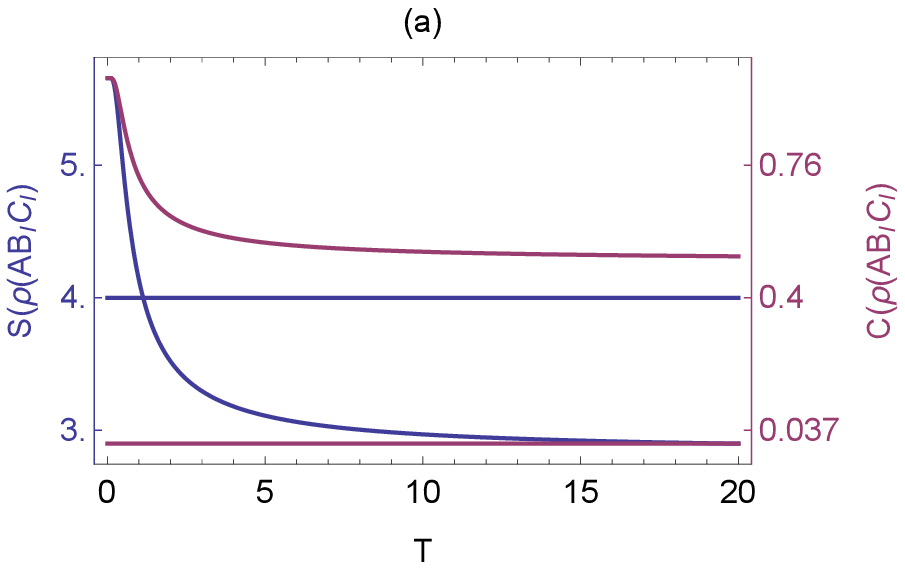}
\label{fig1a}
\end{minipage}%
\begin{minipage}[t]{0.5\linewidth}
\centering
\includegraphics[width=3.0in,height=5.2cm]{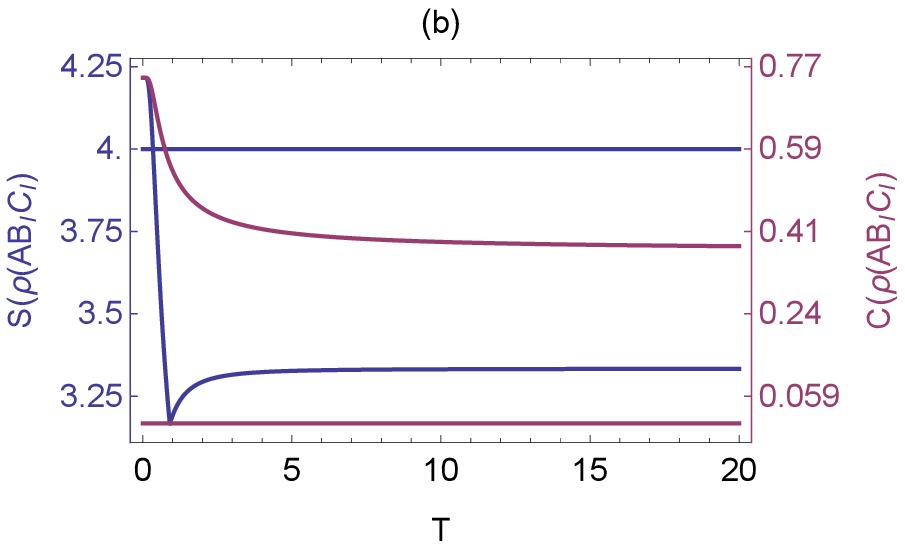}
\label{fig1c}
\end{minipage}%

\begin{minipage}[t]{0.5\linewidth}
\centering
\includegraphics[width=3.0in,height=5.2cm]{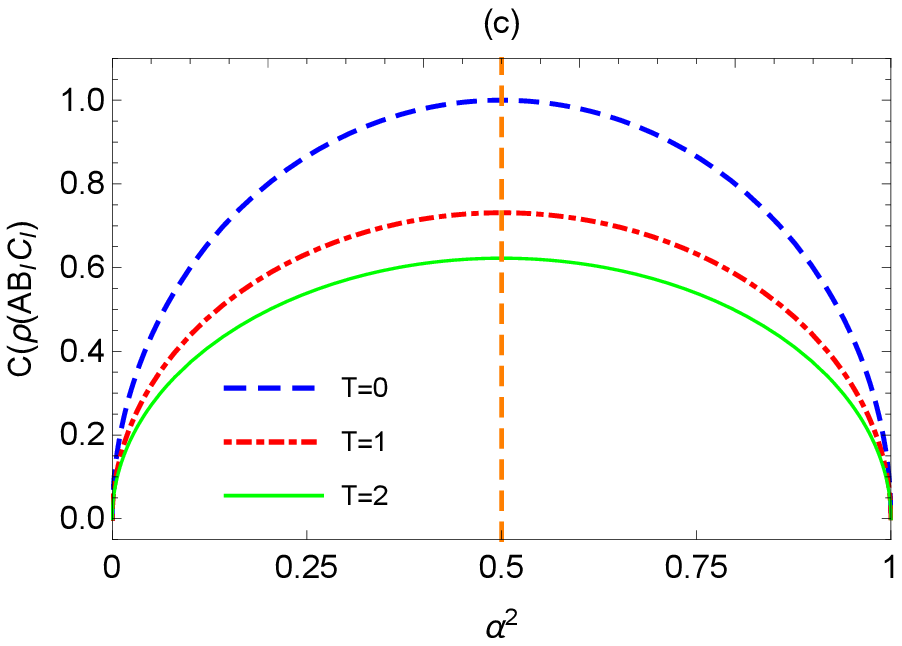}
\label{fig1c}
\end{minipage}%
\begin{minipage}[t]{0.5\linewidth}
\centering
\includegraphics[width=3.0in,height=5.2cm]{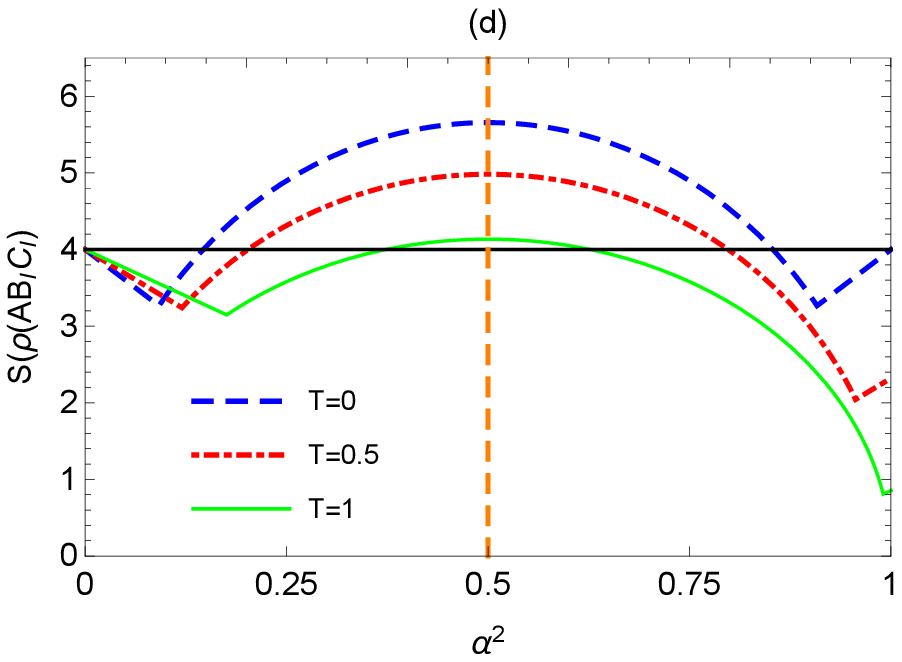}
\label{fig1d}
\end{minipage}%
\caption{The Svetlichny value $S(\rho_{AB_IC_I})$ and the GTE $C(\rho_{AB_IC_I})$ as functions of the
Hawking temperature $T$ or the initial parameter $\alpha^{2}$ for $\omega=1$. (a) $\alpha=\frac{1}{\sqrt{2}}$, and (b) $\alpha=\frac{1}{\sqrt{6}}$. }
\label{Fig1}
\end{figure}

In Fig.\ref{Fig1},  we plot the Svetlichny value $S(\rho_{AB_IC_I})$ and the GTE $C(\rho_{AB_IC_I})$ as functions of the
Hawking temperature $T$ for different initial parameter $\alpha$.
We find that $S(\rho_{AB_IC_I})$  is larger than $4$ at first and then smaller than $4$ with the increase of Hawking temperature $T$. The critical Hawking temperature for $S(\rho_{AB_IC_I})=4$ is $T_{c}=-\frac{\omega}{\ln[2\alpha\sqrt{2(1-\alpha^2)}-1]}$.
This implies that the thermal noise introduced by Hawking temperature destroys the physically accessible GTN between Alice, Bob and Charlie, and takes place ``sudden death" at the critical temperature $T_{c}$.
However, the physically accessible GTE $C(\rho_{AB_IC_I})$ is a monotonic decreasing function of $T$, and has the asymptotic value $\alpha\sqrt{1-\alpha^2}$ in the limit of infinite Hawking temperature. The GTE $C(\rho_{AB_IC_I})$ never takes place ``sudden death" in the finite Hawking temperature. These results suggest that the GTE in the initial state of Eq.(\ref{S14}) can be distinguished into two different parts: nonlocal and local. The nonlocal GTE is destroyed completely by the Hawking effect after the temperature $T>T_{c}$, and finally only the local GTE (whole or partial) is preserved.
At this stage, no quantum information tasks
based on GTN can work, but tasks based on GTE can still work. In other words, GTE is more suitable for relativistic quantum information tasks than GTN.

By comparing Fig.\ref{Fig1}(a) and (b), we find that both the critical temperature $T_{c}$ for the sudden death of GTN and the asymptotic value of $C(\rho_{AB_IC_I})$ in the infinite temperature depend on the initial GTE in Eq.(\ref{S14}), i.e., parameter $\alpha$. The more the initial GTE is, the longer for the death time of GTN is (i.e., the larger the $T_{c}$ is ), and the larger for the asymptotic value of $C(\rho_{AB_IC_I})$ is. In Fig.\ref{Fig1}(a)($\alpha=\frac{1}{\sqrt{2}}$), the critical Hawking temperature (about 1.13) and the asymptotic value (0.5) of $C(\rho_{AB_IC_I})$ are the maximal. For $\alpha=\frac{1}{\sqrt{6}}$ (Fig.\ref{Fig1}(b)), they reduce to about 0.34 and 0.37 respectively.

To further inspect the behaviors of GTN and GTE of the tripartite subsystem $AB_IC_I$ in the regions of the initial parameter $\alpha\in(0,1/\sqrt{2})$ and $\alpha\in(1/\sqrt{2},1)$, we plot Fig.\ref{Fig1}(c) and (d). We can see from Fig.\ref{Fig1}(c) that the GTE $C(\rho_{AB_IC_I})$ is symmetrical with respect to $\alpha^{2}=1/2$ in the parameter region $\alpha^{2}\in[0,1]$. In fact, this symmetry can also be verified via the fact that Eq.(\ref{S18}) is covariant under the exchange $\alpha^{2}\longleftrightarrow 1-\alpha^{2}$. From Fig.\ref{Fig1}(d), we see that the Svetlichny value $S(\rho_{AB_IC_I})$ is obviously asymmetrical with respect to $\alpha^{2}=1/2$. However, the GTN (i.e., the part of $S(\rho_{AB_IC_I})\geq 4$) is symmetrical with respect to $\alpha^{2}=1/2$. This symmetry also can be verified analytically via Eq.(\ref{S17}). Setting $\cos^{2}\eta=(1+e^{-\frac{\omega}{T}})^{-1}$, $\sin^{2}\eta=(1+e^{\frac{\omega}{T}})^{-1}$ and $\cos^{2}\zeta=\alpha^{2}$, $\sin^{2}\zeta=1-\alpha^{2}$, we have $(e^{\frac{\omega}{T}}+1)^{-2}(e^{\frac{\omega}{T}}-1)^{2}=[(1+e^{-\frac{\omega}{T}})^{-1}-(1+e^{\frac{\omega}{T}})^{-1}]^{2}=[\cos^{2}\eta-\sin^{2}\eta]^{2}=\cos^{2}(2\eta)$. Thus the second term in the right hand side of Eq.(\ref{S17}) becomes as $4|\cos^{2}\zeta\cos^{2}(2\eta)+\sin^{2}\zeta|\leq 4$. This means that, in the inspection of the GTN of the tripartite subsystem  $AB_IC_I$, the second term in the right hand side of Eq.(\ref{S17}) can be ignored and only the first term need to be considered. Therefore, the GTN is symmetrical with respect to $\alpha^{2}=1/2$. Note that both the GTE and GTN in Fig.\ref{Fig1}(c) and (d) decrease when Hawking temperature increases, which are consistent with the results from Fig.\ref{Fig1}(a) and (b).
Naturally, in the region $\alpha\in(0,1/\sqrt{2})$,  GTN and GTE  change slowly with the increase of the $\alpha$; in the region $\alpha\in(1/\sqrt{2},1)$,  GTN and GTE  change steeply with the increase of the $\alpha$.

We can also make the similar discussions for other tripartite subsystems. Tracing over the modes $B_{II}$ and $C_{I}$ on the state $\Psi_{AB_IB_{II}C_IC_{II}}$, we obtain the reduced density operator $\rho_{AB_{I}C_{II}}$ as
\begin{eqnarray}\label{S26}
 \rho_{AB_IC_{II}}= \left(\!\!\begin{array}{cccccccc}
n_{1} & 0 & 0 & 0 & 0 & 0 & 0 & 0\\
0 & n_{2} &0  &0  & 0 & 0 & c_{2} & 0\\
0 & 0 & n_{3} & 0 & 0 &  0 &0  & 0\\
0 & 0 & 0 & n_{4} & 0 & 0 & 0 &0 \\
0 & 0 & 0 & 0 & 0 &0 & 0 &0 \\
0 & 0 & 0 & 0 & 0 & 0 & 0 & 0\\
0 & c_{2} & 0 & 0 & 0 & 0 & m_{2} &0 \\
0 & 0 & 0 & 0 & 0 & 0 & 0 & 0
\end{array}\!\!\right),
\end{eqnarray}
with the matrix elements given by
\begin{eqnarray}
\nonumber n_1&=&\alpha^2(e^{-\frac{\omega}{T}}+1)^{-2}, n_2=n_3=\alpha^2 (e^{\frac{\omega}{T}}+e^{-\frac{\omega}{T}}+2)^{-1}\\
a_4&=&\alpha^2(e^{\frac{\omega}{T}}+1)^{-2}, m_2=1-\alpha^2, c_2=\alpha\sqrt{1-\alpha^2}(e^{\frac{\omega}{T}}+e^{-\frac{\omega}{T}}+2)^{-\frac{1}{2}} \nonumber.
\end{eqnarray}
The Svetlichny value and the GTE for this state are given by
\begin{eqnarray}\label{S27}
\nonumber S(\rho_{AB_IC_{II}})=\max \{8\alpha\sqrt{2(1-\alpha^2)}(e^{\frac{\omega}{T}}+e^{-\frac{\omega}{T}}+2)^{-\frac{1}{2}},\\
4|\alpha^2(e^{\frac{\omega}{T}}+1)^{-2}(e^{\frac{\omega}{T}}-1)^{2}-\alpha^2+1|\}.
\end{eqnarray}
and
\begin{eqnarray}\label{S28}
C(\rho_{AB_IC_{II}})=2\alpha\sqrt{1-\alpha^2}(e^{\frac{\omega}{T}}+e^{-\frac{\omega}{T}}+2)^{-\frac{1}{2}},
\end{eqnarray}
respectively. According to the exchange symmetry for Bob and Charlie, we can get
\begin{eqnarray}\label{S29}
S(\rho_{AB_{II}C_{I}})=S(\rho_{AB_IC_{II}}), C(\rho_{AB_{II}C_{I}})=C(\rho_{AB_IC_{II}}).
\end{eqnarray}
Thus the analysis for tripartite system $AB_{II}C_{I}$ is the same as for tripartite system $AB_{I}C_{II}$.

\begin{figure}
\begin{minipage}[t]{0.5\linewidth}
\centering
\includegraphics[width=3.0in,height=5.2cm]{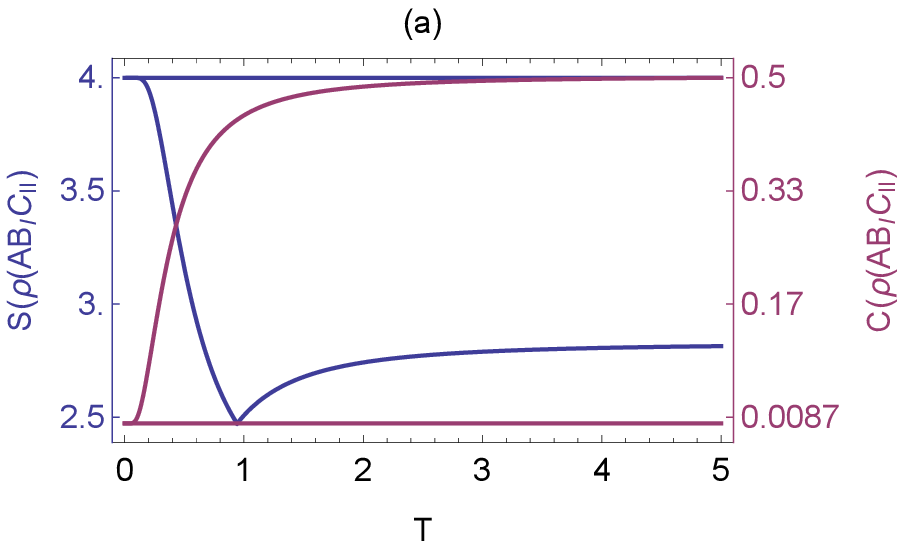}
\label{fig1a}
\end{minipage}%
\begin{minipage}[t]{0.5\linewidth}
\centering
\includegraphics[width=3.0in,height=5.2cm]{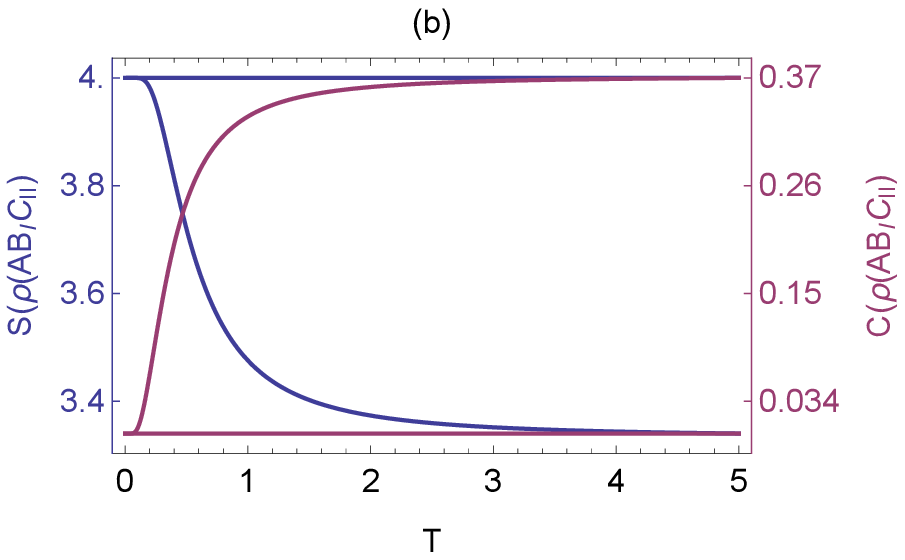}
\label{fig1c}
\end{minipage}%
\caption{The Svetlichny value $S(\rho_{AB_{I}C_{II}})$ and GTE $C(\rho_{AB_{I}C_{II}})$ as functions of the Hawking temperature $T$ for $\omega=1$. (a) $\alpha=\frac{1}{\sqrt{2}}$ and (b) $\alpha=\frac{1}{\sqrt{6}}$. }
\label{Fig2}
\end{figure}

In Fig.\ref{Fig2},  we plot the Svetlichny value $S(\rho_{AB_{I}C_{II}})$ and the GTE $C(\rho_{AB_{I}C_{II}})$ as functions of the Hawking temperature $T$ for different initial parameter $\alpha$.
It is shown that $C(\rho_{AB_{I}C_{II}})$ increases from zero and approaches to the asymptotic value $\alpha\sqrt{1-\alpha^2}$ in the infinite Hawking temperature. This means that the Hawking effect can generate physically inaccessible GTE between modes $A$, $B_{I}$ and $C_{II}$, even though they are separated by the event horizon of black hole. Physically, it can be regarded as a kind of entanglement transfer: Initially, there is GTE between modes $A$, $B_{I}$ and $C_{I}$. Lately, the Hawking effect produce entanglement between the modes $C_{I}$ and $C_{II}$, which is equivalent to an interaction between modes $C_{I}$ and $C_{II}$. This interaction transfers some information from mode $C_{I}$ to mode $C_{II}$. Therefore, the GTE between the modes $A$, $B_{I}$ and $C_{II}$ is established. Comparing Fig.\ref{Fig2} (a) and (b), we find that the produced GTE $C(\rho_{AB_{I}C_{II}})$ depends on the initial accessible GTE in Eq.(\ref{S14}). Under given Hawking temperature, more initially accessible GTE will produce more $C(\rho_{AB_{I}C_{II}})$. In the limit of infinite Hawking temperature, the asymptotic value of $C(\rho_{AB_{I}C_{II}})$ is 0.5 in Fig.\ref{Fig2}(a) and 0.37 in Fig.\ref{Fig2}(b) respectively. The figure shows that $S(\rho_{AB_IC_{II}})$ is always smaller than $4$ for any $T$, thus the physically inaccessible GTN between modes $A$, $B_{I}$ and $C_{II}$ cannot be produced. It also means that the produced inaccessible GTE $C(\rho_{AB_{I}C_{II}})$ is local. Similar analysis is also valid for the tripartite system of modes $A$, $B_{II}$ and $C_{I}$.
The different behaviors between $S(\rho_{AB_{I}C_{II}})$ and $C(\rho_{AB_{I}C_{II}})$ under the Hawking effect suggest that the information flows of different quantum resources inside and outside the event horizon of a black hole are completely different: The entanglement can pass through the event horizon of black hole, while the nonlocality cannot.

Now, we discuss the physically inaccessible GTN and GTE between the modes $A$, $B_{II}$ and $C_{II}$. Tracing over the modes $B_I$ and $C_I$ on the state $\Psi_{AB_IB_{II}C_IC_{II}}$, we obtain the reduced density operator $\rho_{AB_{II}C_{II}}$ as
\begin{eqnarray}\label{S23}
\rho_{AB_{II}C_{II}}= \left(\!\!\begin{array}{cccccccc}
n_{1} & 0 & 0 & 0 & 0 & 0 & 0 & 0\\
0 & n_{2} &0  &0  & 0 & 0 & 0 & 0\\
0 & 0 & n_{3} & 0 & 0 &  0 &0  & 0\\
0 & 0 & 0 & n_{4} & c_4 & 0 & 0 &0 \\
0 & 0 & 0 & c_4 & m_{4} &0 & 0 &0 \\
0 & 0 & 0 & 0 & 0 & 0 & 0 & 0\\
0 & 0 & 0 & 0 & 0 & 0 & 0 &0 \\
0 & 0 & 0 & 0 & 0 & 0 & 0 &0
\end{array}\!\!\right),
\end{eqnarray}
with the matrix elements given by
\begin{eqnarray}
\nonumber n_1&=&\alpha^2(e^{-\frac{\omega}{T}}+1)^{-2}, n_2=n_3=\alpha^2 (e^{\frac{\omega}{T}}+e^{-\frac{\omega}{T}}+2)^{-1}\\
a_4&=&\alpha^2(e^{\frac{\omega}{T}}+1)^{-2}, m_4=1-\alpha^2, c_4=\alpha\sqrt{1-\alpha^2}(e^{\frac{\omega}{T}}+1)^{-1} \nonumber.
\end{eqnarray}
The corresponding Svetlichny value and GTE read
\begin{eqnarray}\label{S24}
S(\rho_{AB_{II}C_{II}})=\max \{8\alpha\sqrt{2(1-\alpha^2)}(e^{\frac{\omega}{T}}+1)^{-1},
4|\alpha^2(e^{\frac{\omega}{T}}+1)^{-2}(e^{\frac{\omega}{T}}-1)^{2}+\alpha^2-1|\},
\end{eqnarray}
and
\begin{eqnarray}\label{S25}
C(\rho_{AB_{II}C_{II}})=2\alpha\sqrt{1-\alpha^2}(e^{\frac{\omega}{T}}+1)^{-1},
\end{eqnarray}
respectively.

\begin{figure}
\begin{minipage}[t]{0.5\linewidth}
\centering
\includegraphics[width=3.0in,height=5.2cm]{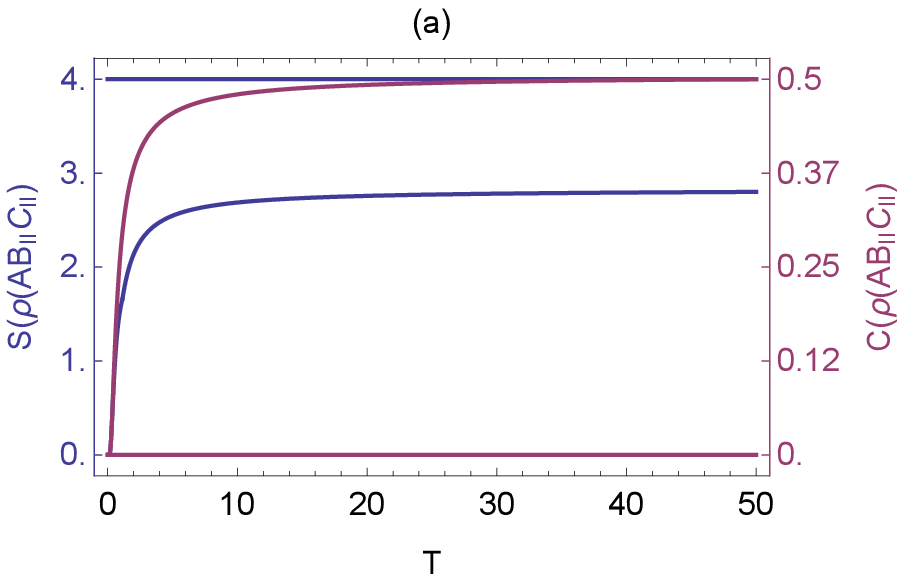}
\label{fig1a}
\end{minipage}%
\begin{minipage}[t]{0.5\linewidth}
\centering
\includegraphics[width=3.0in,height=5.2cm]{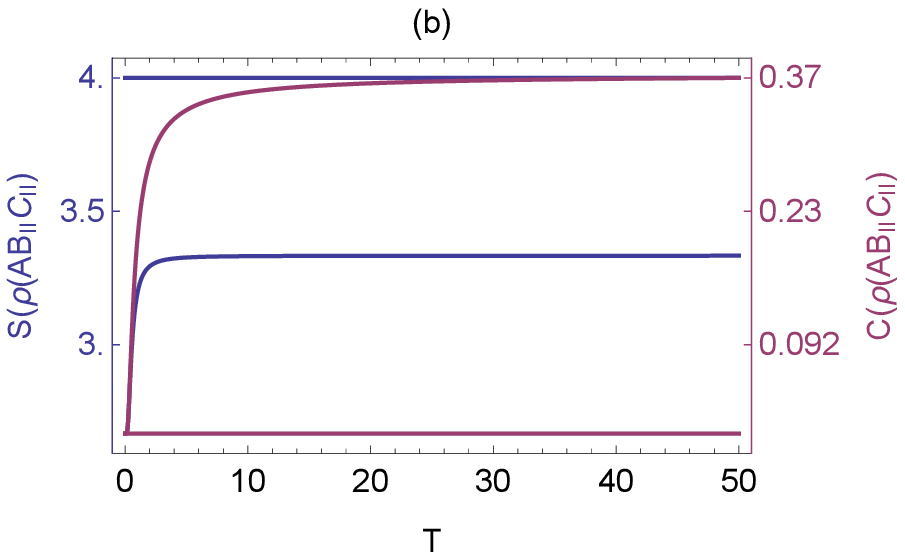}
\label{fig1c}
\end{minipage}%
\caption{ The Svetlichny value $S(\rho_{AB_{II}C_{II}})$ and the GTE $C(\rho_{AB_{II}C_{II}})$ as functions of the
Hawking temperature $T$ for $\omega=1$. (a) $\alpha=\frac{1}{\sqrt{2}}$ and (b) $\alpha=\frac{1}{\sqrt{6}}$. }
\label{Fig3}
\end{figure}

In Fig.\ref{Fig3},  we plot the Svetlichny value $S(\rho_{AB_{II}C_{II}})$ and the GTE $C(\rho_{AB_{II}C_{II}})$ as functions of the
Hawking temperature $T$ for different initial parameter $\alpha$.
We find the similar result as observed in the tripartite system of $AB_{I}C_{II}$: The Hawking effect can generate physically inaccessible GTE between the modes $A$, $B_{II}$ and $C_{II}$, but cannot generate the physically inaccessible GTN, i.e., entanglement can pass through the event horizon of black hole, and nonlocality cannot. The more the initial GTE in Eq.(\ref{S14}) is, the more the produced $C(\rho_{AB_{II}C_{II}})$ is. The produced GTE has the asymptotic value $C(\rho_{AB_{II}C_{II}})=\alpha\sqrt{1-\alpha^2}$ in the limit of infinite Hawking temperature, which is 0.5 for Fig.\ref{Fig3}(a) and 0.37 for Fig.\ref{Fig3}(b) respectively. The mechanism for the production of this GTE is also the result of entanglement transfer.

Finally, we investigate the physically inaccessible GTN and GTE for the system $\rho_{AB_{I}B_{II}}$. In the bases $|000\rangle$, $|100\rangle$, $|010\rangle$, $|001\rangle$, $|101\rangle$, $|111\rangle$, $|110\rangle$, and $|011\rangle$ for $A$, $B_{I}$ and $B_{II}$, the density operator $\rho_{AB_{I}B_{II}}$ has its matrix expression
\begin{eqnarray}\label{S19}
\rho_{AB_{I}B_{II}}= \left(\!\!\begin{array}{cccccccc}
n_{1} & 0 & 0 & 0 & 0 & 0 & 0 & c_1\\
0 & 0 &0  &0  & 0 & 0 & 0 & 0\\
0 & 0 & 0 & 0 & 0 &  0 &0  & 0\\
0 & 0 & 0 & 0 & 0 & 0 & 0 &0 \\
0 & 0 & 0 & 0 & 0 &0 & 0 &0 \\
0 & 0 & 0 & 0 & 0 & 0 & 0 & 0\\
0 & 0 & 0 & 0 & 0 & 0 & m_2 &0 \\
c_1 & 0 & 0 & 0 & 0 & 0 & 0 &m_1
\end{array}\!\!\right),
\end{eqnarray}
with the matrix elements given by
\begin{eqnarray}
\nonumber n_1&=&\alpha^2(e^{-\frac{\omega}{T}}+1)^{-1}, c_1=\alpha^2 (e^{\frac{\omega}{T}}+e^{-\frac{\omega}{T}}+2)^{-\frac{1}{2}}\\
m_1&=&\alpha^2(e^{\frac{\omega}{T}}+1)^{-1}, m_2=1-\alpha^2 \nonumber.
\end{eqnarray}
The corresponding Svetlichny value and the GTE read
\begin{eqnarray}\label{S20}
S(\rho_{AB_IB_{II}})=\max \{8\sqrt{2}\alpha^2(e^{\frac{\omega}{T}}+e^{-\frac{\omega}{T}}+2)^{-\frac{1}{2}},
4|1-2\alpha^2(e^{\frac{\omega}{T}}+1)^{-1}|\},
\end{eqnarray}
and
\begin{eqnarray}\label{S21}
C(\rho_{AB_IB_{II}})=2\alpha^2(e^{\frac{\omega}{T}}+e^{-\frac{\omega}{T}}+2)^{-\frac{1}{2}},
\end{eqnarray}
respectively. According to the exchange symmetry for Bob and Charlie, we can get the Svetlichny value and GTE between modes $A$, $C_I$ and $C_{II}$,
\begin{eqnarray}\label{S22}
S(\rho_{AC_IC_{II}})=S(\rho_{AB_IB_{II}}), C(\rho_{AC_IC_{II}})=C(\rho_{AB_IB_{II}}).
\end{eqnarray}
Thus, we just need to analyze the GTN and GTE for the tripartite system $AB_IB_{II}$.

\begin{figure}
\begin{minipage}[t]{0.5\linewidth}
\centering
\includegraphics[width=3.0in,height=5.2cm]{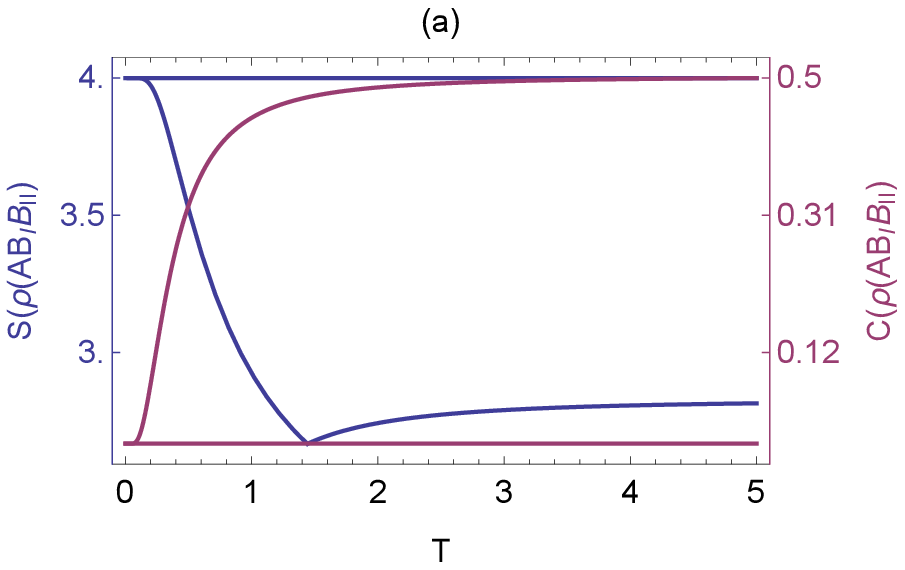}
\label{fig1a}
\end{minipage}%
\begin{minipage}[t]{0.5\linewidth}
\centering
\includegraphics[width=3.0in,height=5.2cm]{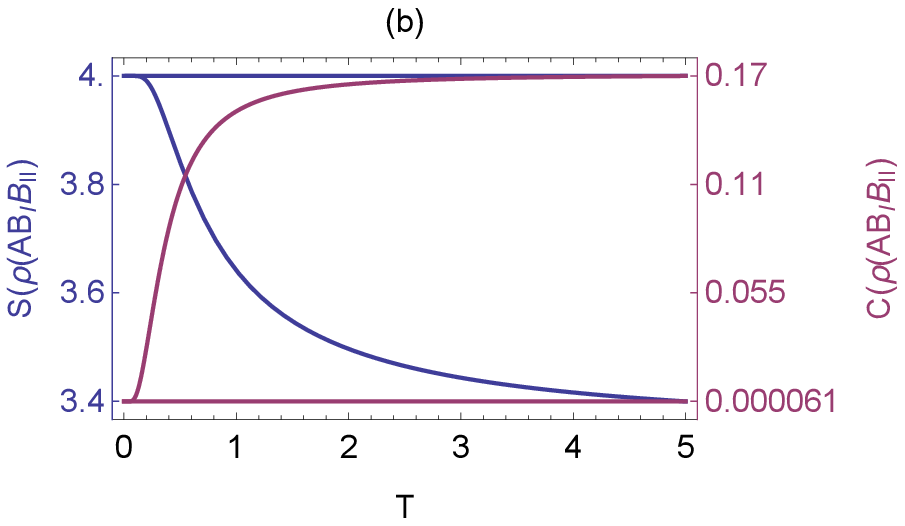}
\label{fig1c}
\end{minipage}%

\begin{minipage}[t]{0.5\linewidth}
\centering
\includegraphics[width=3.0in,height=5.2cm]{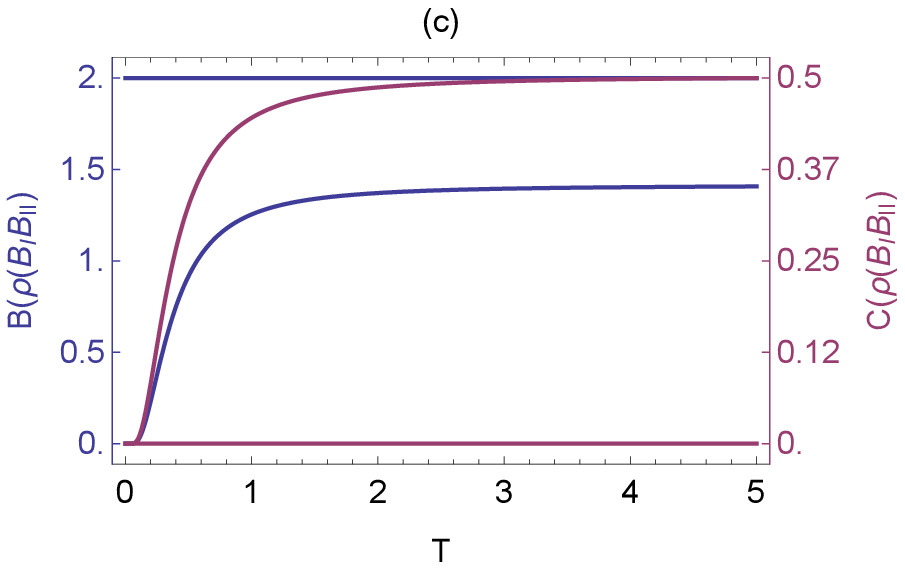}
\label{fig1a}
\end{minipage}%
\begin{minipage}[t]{0.5\linewidth}
\centering
\includegraphics[width=3.0in,height=5.2cm]{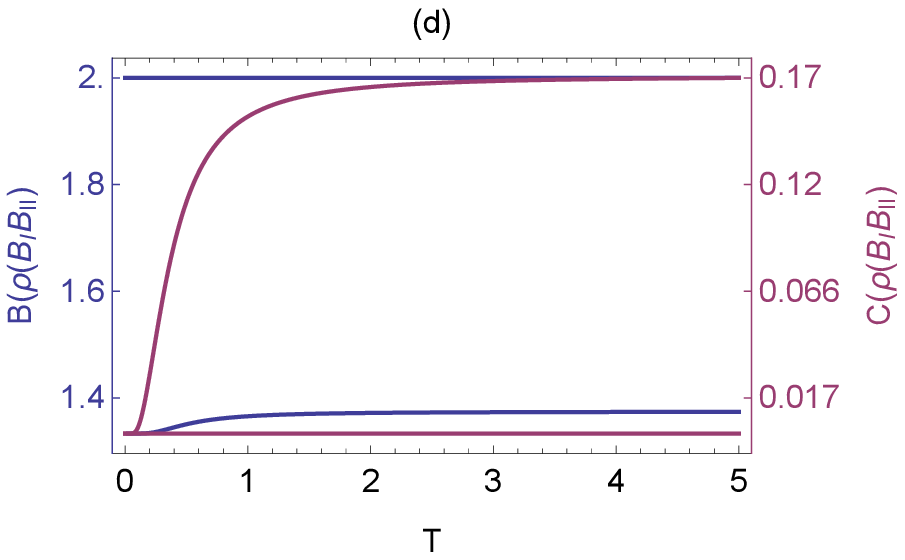}
\label{fig1c}
\end{minipage}%
\caption{ The Svetlichny value $S(\rho_{AB_IB_{II}})$, the GTE $C(\rho_{AB_IB_{II}})$, the maximal Bell signal $\mathcal{B}(\rho_{B_IB_{II}})$, and the bipartite concurrence $C(\rho_{B_IB_{II}})$ as functions of the Hawking temperature $T$ for $\omega=1$. (a) and (c) $\alpha=\frac{1}{\sqrt{2}}$, and (b) and (d) $\alpha=\frac{1}{\sqrt{6}}$. }
\label{Fig4}
\end{figure}

In Fig.\ref{Fig4} (a) and (b),  we plot  the Svetlichny value $S(\rho_{AB_IB_{II}})$ and GTE $C(\rho_{AB_IB_{II}})$ between Alice, Bob and anti-Bob as functions of the
Hawking temperature $T$ for different initial parameter $\alpha$. It is shown that Hawking effect can also produce physically inaccessible GTE between modes $A$, $B_I$ and $B_{II}$, but cannot produce the physically inaccessible GTN between them. The produced $C(\rho_{AB_IB_{II}})$ increases monotonically from zero and reaches the asymptotic value $C(\rho_{AB_IB_{II}})=\alpha^2$ for $T\rightarrow \infty$. Comparing Fig.\ref{Fig4} (a) and (b), we find that $C(\rho_{AB_IB_{II}})$ also depends on the initially accessible GTE in Eq.(\ref{S14}). Under given Hawking temperature, more initially accessible GTE can produce more $C(\rho_{AB_IB_{II}})$.

Besides GTN and GTE for tripartite systems, we can also study the bipartite nonlocality and entanglement for the considered system under the influence of Hawking effect. The bipartite nonlocality and entanglement may be described by the CHSH inequality and concurrence respectively for qubit systems. We review these concepts in the Appendix A and present all the pairwise bipartite nonlocality and entanglement for our considered system in the appendix B.
In Fig.\ref{Fig4} (c)-(d),  we plot the maximal Bell signal $\mathcal{B}(\rho_{B_IB_{II}})$ and bipartite concurrence $C(\rho_{B_IB_{II}})$ between the modes $B_I$ and $B_{II}$ as functions of the
Hawking temperature $T$ for different initial parameter $\alpha$. We find that $\mathcal{B}(\rho_{B_{I}B_{II}})$ is always less than $2$, meaning that
Hawking effect cannot generate Bell nonlocality between Bob and anti-Bob. However, Hawking effect can generate entanglement between Bob and anti-Bob, i.e., entanglement $C(\rho_{B_IB_{II}})$ can pass through the event horizon of black hole. Actually, this is just the physical nature of the Hawking radiation--produce entangled pairs of particle and antiparticle between the causally disconnected regions. The same analysis is also valid for the reduced state $\rho_{C_{I}C_{II}}$. In fact, from the calculation in Appendix B, we can find that all the reduced bipartite subsystems in state $\Psi_{AB_IB_{II}C_IC_{II}}$ have no nonlocality, meaning that nonlocality cannot pass through the event horizon of black hole. Except for $C(\rho_{B_IB_{II}})$ and $C(\rho_{C_IC_{II}})$, there is no pairs of entanglement in any other reduced bipartite subsystems. This result is easy to understand: Initially, there is no entanglement between modes $A$ and $B$, modes $A$ and $C$, modes $B$ and $C$. The Hawking radiation is essentially a local operation on $B_{I}B_{II}$ or $C_{I}C_{II}$, it of course can only produce quantum entanglement between modes $B_{I}$ and $B_{II}$ or $C_{I}$ and $C_{II}$, and cannot produce any other kinds of bipartite entanglement.

From the discussions of Figs.\ref{Fig1}-\ref{Fig4}, we find that the physically accessible
GTN, i.e., $S(\rho_{AB_IC_I})-4$ reduces with Hawking temperature and suffers from a ``sudden death" at some critical Hawking temperature, but the physically inaccessible GTN is never generated by the Hawking effect, i.e., $S(\rho_{AB_IC_{II}})$, $S(\rho_{AB_{II}C_I})$, $S(\rho_{AB_{II}C_{II}})$, $S(\rho_{AB_IB_{II}})$, and $S(\rho_{AC_IC_{II}})$ are all less than 4. This result suggests that the GTN can not be redistributed. However, the GTE behaves differently.
When the physically accessible GTE reduces with Hawking temperature, at the same time, the physically inaccessible
GTE is generated. This result implies that the GTE may be redistributable through the Hawking effect. To manifest this inference, we try to find some monogamy relation for the GTE.
Through careful inspection, we find three monogamy relations between the physically accessible GTE and the physically inaccessible GTE,
\begin{eqnarray}\label{S30}
C(\rho_{AB_{I}C_{I}})+C(\rho_{AB_{II}C_{II}})
=2\alpha\sqrt{1-\alpha^2},
\end{eqnarray}
\begin{eqnarray}\label{S31}
C(\rho_{AB_{I}C_{I}})^2+C(\rho_{AB_{II}C_{I}})^2+C(\rho_{AB_{I}C_{II}})^2+C(\rho_{AB_{II}C_{II}})^2
=4\alpha^2(1-\alpha^2),
\end{eqnarray}
\begin{eqnarray}\label{S32}
\alpha^2[C(\rho_{AB_{I}C_{I}})^2+C(\rho_{AB_{II}C_{II}})^2]+(1-\alpha^2)[C(\rho_{AB_{I}B_{II}})^2+C(\rho_{AC_{I}C_{II}})^2]
=4\alpha^2(1-\alpha^2),
\end{eqnarray}
where $2\alpha\sqrt{1-\alpha^2}$  is the initial GTE in state of Eq.(\ref{S14}). These monogamy relations reflect the restrictions in the redistribution of entanglement from physically accessible to physically inaccessible patterns. Especially the Eq.(\ref{S30}) shows that the total sum of the physically accessible GTE $C(\rho_{AB_{I}C_{I}})$ and the physically inaccessible GTE $C(\rho_{AB_{II}C_{II}})$ is equal to the initial GTE. The monogamy relations are important for understanding the transfer of quantum information in relativistic spacetime.

Besides the above monogamy relations, we also find that the physically accessible entanglement fulfills the following Coffman-Kundu-Wootters monogamy inequality  \begin{eqnarray}\label{S33}
C(\rho_{ijk})^2\geq C(\rho_{ij})^2+C(\rho_{ik})^2,
\end{eqnarray}
where ($i, j, k$) denote all the permutations of the three modes $A, B_{I}, C_{I}$.  These Coffman-Kundu-Wootters monogamy inequalities reflect the distribution of the physically accessible entanglement in the environment of Schwarzschild black hole.

\section{ Conclusions  \label{GSCDGE}}
The effect of Hawking radiation on the GTN and GTE for Dirac fields in Schwarzschild spacetime has been investigated.
It has been shown that Hawking effect degrades both the physically accessible GTN and the physically accessible GTE, where the former takes place ``sudden death" at some critical Hawking temperature, and the latter approaches to the nonzero asymptotic value in the infinite Hawking temperature. This means that on the one hand the surviving physically accessible GTE is not nonlocal, and on the other hand the GTE has more resistance to the Hawking noise than GTN.

Further investigation has demonstrated that the GTN is not redistributable, but GTE can be redistributed through Hawking effect. With the growth of the Hawking temperature, the physically accessible GTN decreases, but no physically inaccessible GTN is generated. The GTE however behaves differently: With the loss of the physically accessible GTE, the physically inaccessible GTE is generated continually through Hawking effect. Further, the physically accessible GTE and the physically inaccessible GTE fulfil some monogamy relations. All these phenomena suggest that the GTE is redistributable. We can regard the redistribution of entanglement as a kind of phenomenon of information tunnelling, i.e. the flow of quantum entanglement can pass through the event horizon of black hole, but the flow of quantum nonlocality can not. This result has been demonstrated also by the pairwise bipartite nonlocality and entanglement in the underlying system considered in this paper.

Note that the authors in reference \cite{L31} studied the tripartite entanglement in environment of Schwarzschild black hole used the measure of $\pi$-tangle.
They found that the physically accessible $\pi$-tangle decreases with Hawking temperature and approaches to a nonzero asymptotic value for the infinite Hawking temperature. Similar Coffman-Kundu-Wootters monogamy inequalities to Eq.(\ref{S33}) for $\pi$-tangle was found. These results agree with ours.

Quantum entanglement and nonlocality are the important manifestations of quantum correlation. They have potential applications in various fields of science.
We expect our research can present helps for these applications and enriches the theory of the relativistic quantum information science.

\begin{acknowledgments}
This work is supported by the National Natural
Science Foundation of China (Grant Nos. 1217050862, 11275064), and the Construct Program of the National Key Discipline.	
\end{acknowledgments}

\appendix
\onecolumngrid

\section{CHSH inequality and concurrence for bipartite systems }
In this appendix, we review the concepts of CHSH inequality and concurrence for bipartite systems. The CHSH inequality is an important tool for judging quantum nonlocality. It is considered to be both necessary and sufficient conditions of the separability for bipartite pure states. The key element for the CHSH inequality is the Bell operator defined by \cite{L50}
\begin{eqnarray}\label{A1}
\mathcal{B}_{CHSH}=\boldsymbol{a}\cdot\boldsymbol{\sigma}\otimes(\boldsymbol{b}+\boldsymbol{b'})
\cdot\boldsymbol{\sigma}+\boldsymbol{a'}\cdot\boldsymbol{\sigma}\otimes(\boldsymbol{b}-\boldsymbol{b'})
\cdot\boldsymbol{\sigma},
\end{eqnarray}
where  $\boldsymbol{a}$, $\boldsymbol{a'}$, $\boldsymbol{b}$ and $\boldsymbol{b'}$
are unit vectors in $\mathbb{R}^3$, and $\boldsymbol{\sigma}=(\sigma_1, \sigma_2, \sigma_3)$ is the vector of Pauli matrices. For any bipartite mixed state $\rho$,
the well-known CHSH inequality can be expressed as
\begin{eqnarray}\label{A2}
\mathcal{B}(\rho)=|{\rm Tr}(\rho\mathcal{B}_{CHSH})|\leq2.
\end{eqnarray}
CHSH inequality holds for any state $\rho$ that admits local hidden variable model, and the violation of it implies the Bell nonlocality of the underlying state.
In practical applications, we need to find the maximal Bell signal $\mathcal{B}(\rho)$, which can be equivalently expressed for two-qubit systems as
\begin{eqnarray}\label{A3}
\mathcal{B}(\rho)=2\sqrt{\max_{i<j}(\mathcal{Z}_i+\mathcal{Z}_j)},
\end{eqnarray}
where $\mathcal{Z}_i$ and $\mathcal{Z}_j$ are the two largest eigenvalues of
$U(\rho)=T^{\rm T}_{\rho}T_{\rho}$, and the correlation matrix is defined by $T=(t_{ij})$ with $t_{ij}={\rm Tr}[\rho\sigma_i\otimes\sigma_j]$.
Bell nonlocality can be witnessed by the maximum violation of CHSH inequality.

For two-qubit X-state,
\begin{eqnarray}\label{A4}
\rho^X= \left(\!\!\begin{array}{cccccccc}
\rho_{11} & 0 & 0 & \rho_{14} \\
0 & \rho_{22} &\rho_{23} &0 \\
0 & \rho_{23} & \rho_{33} & 0\\
\rho_{14} & 0 & 0 & \rho_{44}
\end{array}\!\!\right),
\end{eqnarray}
with all elements $\rho_{ij}$ being real, the three eigenvalues corresponding to the matrix $U(\rho)=T^{\rm T}_{\rho}T_{\rho}$ are,
\begin{eqnarray}\label{A5}
\mathcal{Z}_1=4(|\rho_{14}|+|\rho_{23}|)^2, \mathcal{Z}_2=4(|\rho_{14}|-|\rho_{23}|)^2,
\mathcal{Z}_3=(|\rho_{11}|-|\rho_{22}|-|\rho_{33}|+|\rho_{44}|)^2.
\end{eqnarray}
As $\mathcal{Z}_1$ is greater than $\mathcal{Z}_2$, so the maximal Bell signal reads
\begin{eqnarray}\label{A6}
\mathcal{B}(\rho^X)=\max\{\mathcal{B}_1, \mathcal{B}_2\},
\end{eqnarray}
with $\mathcal{B}_1=2\sqrt{\mathcal{Z}_1+\mathcal{Z}_2}$ and $\mathcal{B}_2=2\sqrt{\mathcal{Z}_1+\mathcal{Z}_3}$ \cite{L51,L52}.

In addition, the concurrence for quantifying the entanglement of the two-qubit
X-state of Eq.(\ref{A4}) can be calculated through the expression,
\begin{eqnarray}\label{A7}
C(\rho^X)=2\max\{|\rho_{14}|-\sqrt{\rho_{22}\rho_{33}}, |\rho_{23}|-\sqrt{\rho_{11}\rho_{44}}\},
\end{eqnarray}
where $\rho_{ij}$ is the element of density matrix $\rho^X$ \cite{L45}.

\section{Pairwise quantum nonlocality and entanglement in state $\Psi_{AB_IB_{II}C_IC_{II}}$ }
We now study the pairwise bipartite quantum nonlocality and entanglement in the state $\Psi_{AB_IB_{II}C_IC_{II}}$.
By tracing over the irrelevant modes on state $\Psi_{AB_IB_{II}C_IC_{II}}$, we obtain all the pairwise density operators as
\begin{eqnarray}\label{B1}
\rho_{B_{I}B_{II}}= \left(\!\!\begin{array}{cccccccc}
\frac{\alpha^2}{(e^{-\frac{\omega}{T}}+1)} & 0 & 0 & \frac{\alpha^2}{\sqrt{e^{\frac{\omega}{T}}+e^{-\frac{\omega}{T}}+2}} \\
0 & 0 &0 &0 \\
0 & 0 & 1-\alpha^2 & 0\\
\frac{\alpha^2}{\sqrt{e^{\frac{\omega}{T}}+e^{-\frac{\omega}{T}}+2}} & 0 & 0 & \frac{\alpha^2}{(e^{\frac{\omega}{T}}+1)}
\end{array}\!\!\right),
\end{eqnarray}

\begin{eqnarray}\label{B2}
\rho_{AB_{I}}=\rho_{AC_{I}}= \left(\!\!\begin{array}{cccccccc}
\frac{\alpha^2}{(e^{-\frac{\omega}{T}}+1)} & 0 & 0 & 0 \\
0 & \frac{\alpha^2}{(e^{\frac{\omega}{T}}+1)} &0 &0 \\
0 & 0 & 0 & 0\\
0 & 0 & 0 & 1-\alpha^2
\end{array}\!\!\right) ,
\end{eqnarray}

\begin{eqnarray}\label{B3}
\rho_{B_IC_{I}}=\left(\!\!\begin{array}{cccccccc}
\frac{\alpha^2}{(e^{-\frac{\omega}{T}}+1)^{2}} & 0 & 0 & 0 \\
0 & \frac{\alpha^2}{(e^{\frac{\omega}{T}}+e^{-\frac{\omega}{T}}+2)} &0 &0 \\
0 & 0 & \frac{\alpha^2}{(e^{\frac{\omega}{T}}+e^{-\frac{\omega}{T}}+2)} & 0\\
0 & 0 & 0 & \frac{\alpha^2}{(e^{\frac{\omega}{T}}+1)^{2}}+1-\alpha^2,
\end{array}\!\!\right) ,
\end{eqnarray}

\begin{eqnarray}\label{B4}
\rho_{B_{II}C_{II}}=\left(\!\!\begin{array}{cccccccc}
\frac{\alpha^2}{(e^{-\frac{\omega}{T}}+1)^{2}}+1-\alpha^2 & 0 & 0 & 0 \\
0 & \frac{\alpha^2}{(e^{\frac{\omega}{T}}+e^{-\frac{\omega}{T}}+2)} &0 &0 \\
0 & 0 & \frac{\alpha^2}{(e^{\frac{\omega}{T}}+e^{-\frac{\omega}{T}}+2)} & 0\\
0 & 0 & 0 & \frac{\alpha^2}{(e^{\frac{\omega}{T}}+1)^{2}},
\end{array}\!\!\right) ,
\end{eqnarray}

\begin{eqnarray}\label{B5}
\rho_{AB_{II}}=\rho_{AC_{II}}= \left(\!\!\begin{array}{cccccccc}
\frac{\alpha^2}{(e^{-\frac{\omega}{T}}+1)} & 0 & 0 & 0 \\
0 & \frac{\alpha^2}{(e^{\frac{\omega}{T}}+1)} &0 &0 \\
0 & 0 &  1-\alpha^2 & 0\\
0 & 0 & 0 &0
\end{array}\!\!\right) ,
\end{eqnarray}

\begin{eqnarray}\label{B6}
\rho_{B_IC_{II}}=\rho_{B_{II}C_{I}}= \left(\!\!\begin{array}{cccccccc}
\frac{\alpha^2}{(e^{-\frac{\omega}{T}}+1)^{2}} & 0 & 0 & 0 \\
0 & \frac{\alpha^2}{(e^{\frac{\omega}{T}}+e^{-\frac{\omega}{T}}+2)} &0 &0 \\
0 & 0 &  \frac{\alpha^2}{(e^{\frac{\omega}{T}}+e^{-\frac{\omega}{T}}+2)}+1-\alpha^2 & 0\\
0 & 0 & 0 &\frac{\alpha^2}{(e^{\frac{\omega}{T}}+1)^{2}}
\end{array}\!\!\right) .
\end{eqnarray}
Using Eqs.(\ref{A6}) and (\ref{A7}), it is easy to find the maximal Bell signal and concurrence for the corresponding pairwise qubit systems,
\begin{eqnarray}\label{B7}
\mathcal{B}(\rho_{B_{I}B_{II}})=\max\{4\sqrt{2}\alpha^2 (e^{\frac{\omega}{T}}+e^{-\frac{\omega}{T}}+2)^{-\frac{1}{2}},
2\sqrt{(2\alpha^2-1)^2+4\alpha^2 (e^{\frac{\omega}{T}}+e^{-\frac{\omega}{T}}+2)^{-\frac{1}{2}}}\},
\end{eqnarray}
\begin{eqnarray}\label{B8}
&&\mathcal{B}(\rho_{AB_{I}})=\mathcal{B}(\rho_{AC_{I}})=2[\alpha^2(e^{\frac{\omega}{T}}+1)^{-1}(e^{\frac{\omega}{T}}-1)+1-\alpha^2]<2,\\
&&\mathcal{B}(\rho_{B_IC_{I}})=\mathcal{B}(\rho_{B_{II}C_{II}})=2[\alpha^2(e^{\frac{\omega}{T}}+1)^{-2}(e^{\frac{\omega}{T}}-1)^2+1-\alpha^2]<2,\\
&&\mathcal{B}(\rho_{AB_{II}})=\mathcal{B}(\rho_{AC_{II}})=2[\alpha^2(e^{\frac{\omega}{T}}+1)^{-1}(e^{\frac{\omega}{T}}-1)-1+\alpha^2]<2, \\&&\mathcal{B}(\rho_{B_IC_{II}})=\mathcal{B}(\rho_{B_{II}C_{I}})=2[\alpha^2(e^{\frac{\omega}{T}}+1)^{-2}(e^{\frac{\omega}{T}}-1)^2-1+\alpha^2 ]<2,
\end{eqnarray}
and
\begin{eqnarray}\label{B9}
C(\rho_{B_IB_{II}})=\max \{0,2\alpha^2(e^{\frac{\omega}{T}}+e^{-\frac{\omega}{T}}+2)^{-\frac{1}{2}}\},
\end{eqnarray}
\begin{eqnarray}\label{B13}
&&C(\rho_{AB_{I}})=C(\rho_{AC_{I}})=0,\\
&&C(\rho_{B_IC_{I}})=C(\rho_{B_{II}C_{II}})=0,\\
&&C(\rho_{AB_{II}})=C(\rho_{AC_{II}})=0, \\
&&C(\rho_{B_IC_{II}})=C(\rho_{B_{II}C_{I}})=0.
\end{eqnarray}


\end{document}